\documentclass[a4paper,12pt]{article}
\pdfoutput=1
\usepackage{authblk}
\usepackage{blindtext}
\usepackage{soul}
\usepackage[usenames,dvipsnames]{color}

\usepackage{amsmath, amssymb, slashed, epsf, color, graphicx, latexsym}
\usepackage{epsfig}
\usepackage{graphics}
\usepackage{tikz}
\newcommand{\be}{\begin{equation}}
\newcommand{\ee}{\end{equation}}
\newcommand{\bea}{\begin{eqnarray}}
\newcommand{\eea}{\end{eqnarray}}
\def\bse{\begin{subequations}}
\def\ese{\end{subequations}}
\newcommand{\IR}{\mathbb{R}} 
 
\def\IZ{\relax\ifmmode\hbox{Z\kern-.4em Z}\else{Z\kern-.4em Z}\fi}
\newcommand{\non}{\nonumber \\}

\def\del{{\partial}} 
\def\presub{\vspace{.5cm} \noindent}

\def\bi{\begin{itemize}} \def\ei{\end{itemize}}

\def\({\left(} \def\){\right)}
\newcommand{\al}{\alpha}

\newcommand{\ep}{\epsilon}

\newcommand{\la}{\lambda}       

\title{Kite diagram through Symmetries of Feynman Integrals}
\author{Barak Kol and Subhajit Mazumdar \\
{\it The Racah Institute of Physics, Hebrew University,}\\ \it{Jerusalem 91904, Israel}\\
{\tt E-mail:  barak.kol, mazumdar.subhajit@mail.huji.ac.il}
}
\date{}

\begin{document}
\maketitle
%	\begin{center}
%	\textbf{Kite diagram through Symmetries of Feynman Integrals}
%	\end{center}
%	\begin{center}
%	Barak Kol and Subhajit Mazumdar 
%	\end{center}
%	\textit{Racah Institute of Physics, Hebrew University, Jerusalem 91904, Israel}
%	barak.kol, mazumdar.subhajit@mail.huji.ac.il\\
%	
	\begin{center}
	\textbf{Abstract}
\end{center}
	
The Symmetries of Feynman Integrals (SFI) is a method for evaluating Feynman Integrals which exposes a novel continuous group associated with the diagram which depends only on its topology and acts on its parameters. Using this method we study the kite diagram, a two-loop diagram with two external legs, with arbitrary masses and spacetime dimension. Generically, this method reduces a Feynman integral into a line integral over simpler diagrams. We identify a locus in parameter space where the integral further reduces to a mere linear combination of simpler diagrams, thereby maximally generalizing the known massless case.

\newpage
\tableofcontents
\section{Introduction}
 
The Symmetries of Feynman Integrals method \cite{SFI}  considers a Feynman diagram of fixed topology,
\footnote{By diagram topology we mean the standard mathematical definition of a graph.} but varying masses, kinematical invariants and spacetime dimension. Each diagram is associates with a set of differential equations in this parameter space. The equation set defines a continuous symmetry group $G$ which acts on parameter space and foliates it into orbits. This geometry allows to reduce the diagram to its value at some convenient base point within the same orbit plus a line integral over simpler diagrams (with one edge contracted).

The SFI method is related to both the Integration By Parts method \cite{ChetyrkinTkachov1981} as well as to the Differential Equations method \cite{DE1:Kotikov1990, DE2:Remiddi1997, DE3:GehrmannRemiddi1999}, see also the textbook \cite{SmirnovBook2006, SmirnovBook2012}. The new elements include the definitions of the group and its orbits, as well as the reduction to a line integral.

\begin{figure}
\centering \noindent
\includegraphics[width=10cm]{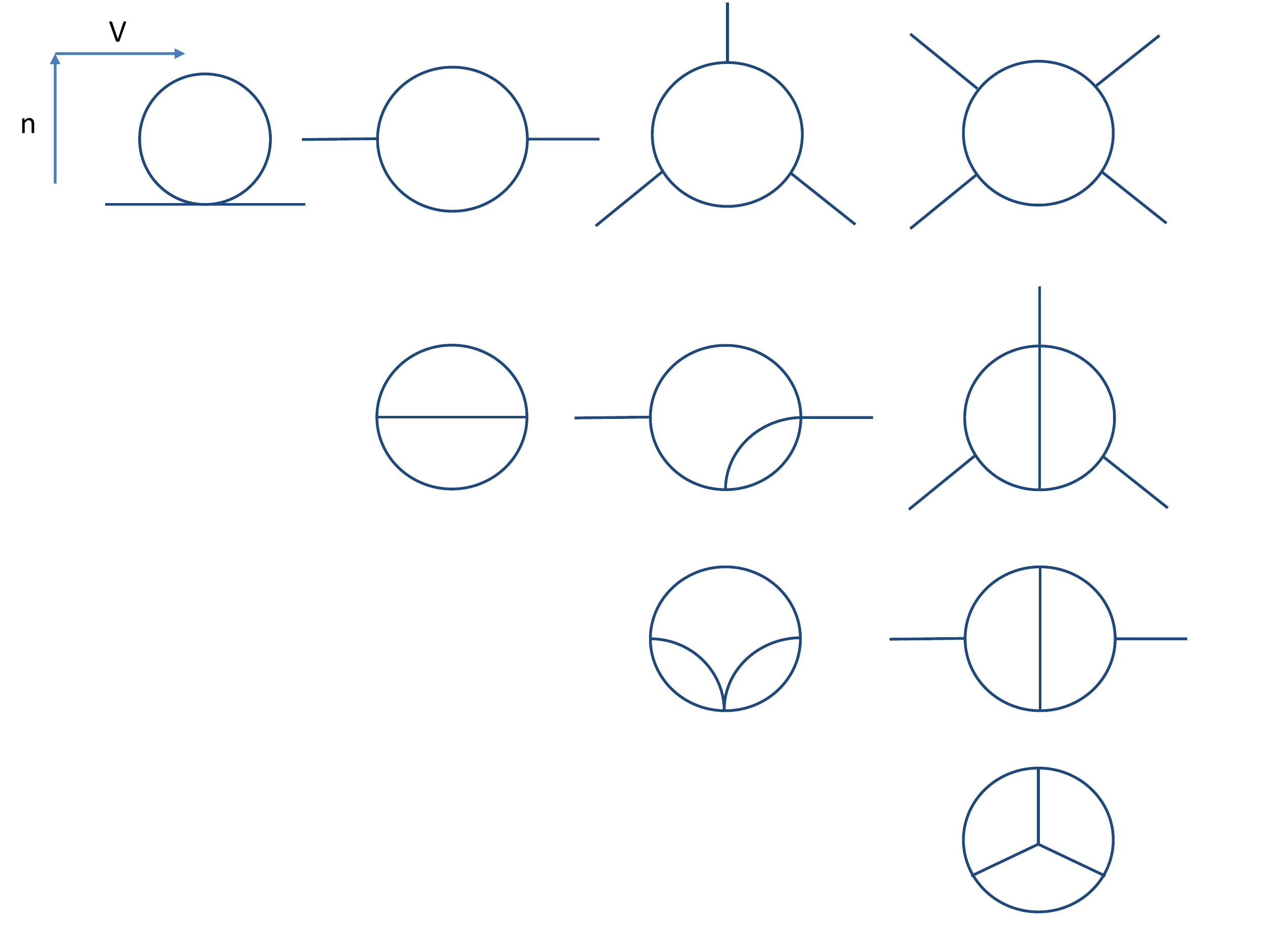} 
\caption[]{Hierarchy of diagrams according to edge contraction.  Each column has diagrams of fixed number of vertices $V=1,2,3,4$. Since contraction reduces $V$ by one the necessary sources for each diagram are always on its left. Each column in ordered according to the number of external legs $n$. The kite is on the rightmost column.

Not all diagrams of given $V,n$ are shown. In particular each diagram can produce others with the same values of $V,n$ by adding propagators between existing vertices.}
 \label{fig:DiagHierarchy} \end{figure}

Since its introduction the SFI method was developed and applied to several diagrams in \cite{locus, bubble, VacSeagull, minors, diam}. The method suggests to partially order all diagrams according to edge contraction as shown in fig. \ref{fig:DiagHierarchy} where the sources for each diagram are in the columns to its left. The tadpole on the leftmost of the figure is the simplest and its evaluation is immediate. The 1-loop propagator diagram, or the bubble, to its right can be evaluated directly through the $\alpha$ variables and was analyzed through SFI in \cite{bubble}. The diagram just below it in the figure, namely the 2-loop vacuum diagram, or ``diameter'' was analyzed through SFI in \cite{diam}.  A 3-loop diagram, the vacuum seagull, is on the third column from the left at the bottom. It was analyzed through SFI in \cite{VacSeagull}, enabling a novel evaluation of a sector with three mass scales.

This paper studies the kite diagram, which is on the rightmost column of the figure, second from bottom. It is the first diagram with four vertices to be analyzed through SFI, so it is the first of its column, see also \cite{tetrahedron}. 

\begin{figure}
\centering \noindent
\includegraphics[width=4.5cm]{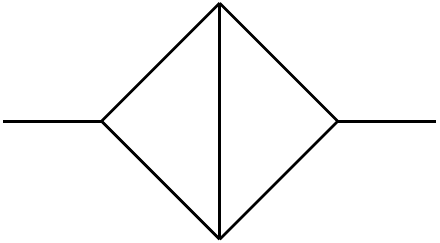} 
\caption[]{The kite diagram drawn in a way which explains its name.}
\label{fig:kite1} \end{figure}

This diagram appeared in the electron propagator renormalization of QED, at least as early as \cite{Sabry1962}, where a single mass scale was studied (however, neither the diagram nor even the term Feynman diagram appear there). The Integration By Parts method \cite{ChetyrkinTkachov1981} allowed a reduction of the massless case to a linear combination of simpler diagrams. QCD required two mass-scales and was studied in 4d in \cite{Broadhurst1990} where the diagram was reduced to a line integral over logarithms and was called by the rather general name ``the master diagram''. \cite{Bauberger:1994hx} presented an expression for the general kite through a dispersion integral.  \cite{Tarasov:1997kx} studied all two-loop diagrams of propagator type and determined the relevant master diagrams which include the kite, but did not study it. \cite{BernDixonKosower2004} encountered the diagram while studying gluon splitting in QCD. It introduced the name ``lizard-eye bubble'' which does not seem to have been adopted by the literature. The 3d massless version of the diagram was found to be essential in the second post-Newtonian approximation (2PN) of the two-body problem in Einstein's gravity \cite{GilmoreRoss2008}. More recently \cite{RemiddiTancredi2016} studied the diagram while applying dispersion relations within the Differential Equations method and encountering the kite with a single mass scale. It introduced the term ``kite diagram'', motivated by a slightly different way of drawing the diagram, see fig. \ref{fig:kite1}. The round version of drawing, e.g. fig.s \ref{fig:DiagHierarchy}-\ref{fig:mars}, could be called also ``the marshmallow diagram''. Finally, \cite{Adams:2016xah, Bogner:2017vim} found that such kite integrals can be expressed in terms of elliptic generalizations of (multiple) polylogarithms.

In this paper we shall study the most general parameters for the kite, with as many as five different masses. We shall ask \bi

\item How big are the $G$-orbits in parameter space? More specifically, what is their co-dimension?

\item What is the locus where the diagram degenerates into a linear combination of simpler ones (rather than a line integral over them)?  This is known as the algebraic locus \cite{locus}. What is the associated solution?

\ei

The paper is organized as follows. Section \ref{sec:eq_set} introduces the diagram and describes the SFI equation set and the associated group. In section \ref{sec:geom} we study the orbit geometry in parameter space, obtain the answer to the first question and find the homogeneous solution.  Section \ref{sec:algeb} answers the second question and finally section \ref{sec:summ} offers a summary and discussion.

%%%%%%%%%%%%%%%%%%%%%%%%%%%%%%%%%%%%%%%%%%
\section{Equation set}
\label{sec:eq_set}

\begin{figure}
\centering \noindent
\includegraphics[width=9cm]{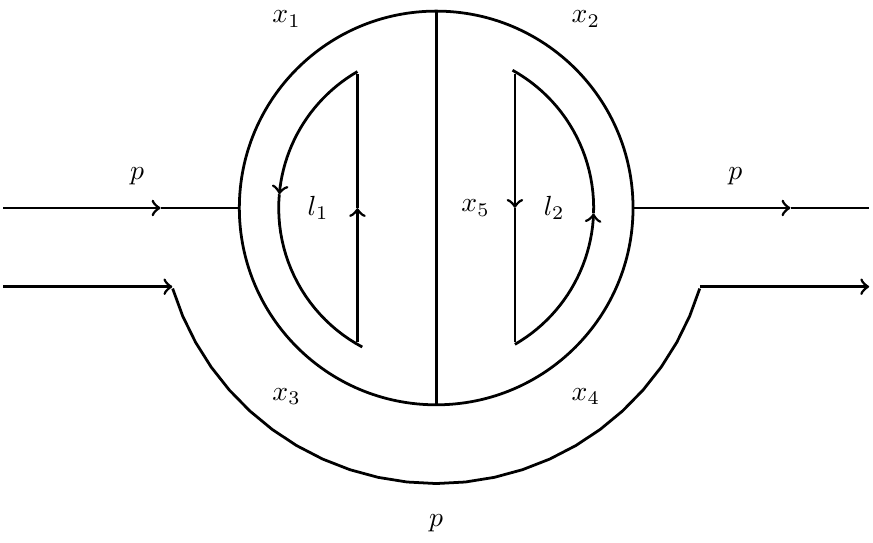} 
\caption[]{The kite diagram with its parameters and a choice of currents.}
\label{fig:mars} \end{figure}

{\bf Definition of diagram and integral}. Consider the kite diagram shown in the fig. \ref{fig:mars}. It has $L=2$ loops, $n=2$ external legs and the associated integral is given by \bea
 &I&(p^2 ; x_{1},x_{2},x_{3},x_{4},x_{5})  = \non
  &=& \int \frac{d^{d}l_{1}\, d^{d}l_{2}}{(l_{1}^2-x_{1})(l_{2}^2-x_{2})((l_{1}+p)^2-x_{3})((p+l_{2})^2-x_{4})((l_{1}-l_{2})^2-x_{5})}
\label{def:I}
\eea
%
%%%%% Kite Figure %%%%%%%%%%%%%%%%%%%%

%%%%%%%%%%%%%%%%%%%%%%%%%%%%%%%%%%%%%%%%

The integral is a function of six parameters: five mass-squares and a single kinematical invariant, namely $p^2$, the square of the incoming momentum. Accordingly, the parameter space $X$ is given by \be
X= \left\{ (x_1,...,x_5,x_6) = (\, (m_1)^2, \dots,(m_5)^2, p^2 ) \right \}
\ee
The figure defines our choice of loop currents $l_{1}$ and $l_{2}$ and the routing of $p$. We consider a general spacetime dimension $d$ and the mass dimension of the integral is $2\, d - 10$.

The discrete symmetry group $\Gamma$, namely the standard symmetries of the Feynman diagram, are given by reflections: either right-left (R) or up-down (U), that is \be
\Gamma = \IZ_2^R \times \IZ_2^U ~.
\label{def:Gamma}
\ee

We will study the diagram through the Symmetries of Feynman Integrals method (SFI) described in \cite{SFI}. Briefly, one varies the integral with respect to infinitesimal re-definitions of loop momenta thereby giving rise to a set of differential equations which the integral satisfies in parameter space $X$.  Let us determine the equation set for $I$ and the associated group $G$.

\presub {\bf The SFI group}. $G$ is known to be a subset of certain triangular matrices \cite{bubble} \be
 G \subseteq T_{L,n-1} \equiv T_{2,1} 
\label{GsubsetT}
 \ee
where $T_{L,n-1}$ represents the block upper triangular matrices such that the first block is of size $L$ and the second
one is $n-1$.

The obstructions for $T_{2,1}$ generators are related to potential numerators of the diagram. The potential numerators are the quotient of the quadratics by the squares (of propagator currents) \bea
Num &=& Qd/Sq = \\
 &=& {\rm Sp} \left\{ l_1^2, l_2^2,l_1 \cdot l_2, p \cdot l_1, p \cdot l_2,p^2 \right\} / {\rm Sp} \left\{ l_1^2, l_2^2, (l_1+p)^2, (l_2 + p)^2, (l_1-l_2)^2, p^2 \right\} = \emptyset \nonumber
\eea
So the kite does not have potential numerators and hence $G$ saturates (\ref{GsubsetT}), namely \be
G = T_{2,1} \equiv
\( \begin{array}{ccc}
 * & * & * \\ 
 * & * & * \\
 0 & 0 & * \\
 \end{array} \)
\ee
and the number of equations is \be
{\rm dim}(T_{2,1})=7 ~.
\label{dimG}
\ee
More precisely the Lie algebra is $T_{2,1}$ and the group $G$ consists of invertible upper triangular matrices.

\presub {\bf The SFI equation set}. As a basis for the space of generators we choose  \be
 \( \begin{array}{c}
	E^1 \\ % old E1
	E^2 \\ % old E5 + E1
	E^3 \\ % old E1-E3
	E^4 \\ % old E2
	E^5 \\ % old E6+ E2
	E^6 \\ % old E2-E4
	E^7 \\  % old E1 + E2 + E7
	\end{array}
	\) =
  \( \begin{array}{c}
	l_1\, \del_l^1 \\
	-(l_1-p) \del_l^1 \\
	l_1 (\del_l^2 - \del_l^1) \\
	l_2\, \del_l^2 \\
	-(l_2-p) \del_l^2 \\
	l_2 (\del_l^1 - \del_l^2) \\
	l_1\, \del_l^1 + l_2 \del_l^2 + p \del_p \\
	\end{array}
	\)
\label{def:basis}
\ee
where $\del_l^a \equiv \del/\del_{l_a}$.  

The equations are given by the usual SFI form \be
 c^a\,  I+ Tx^a_j\, \partial^j\, I + J^a=0
 \label{eq_set}
\ee
where $c^a, Tx^a_j$ and $J^a$ shall be defined immediately within the above-mentioned basis. The vector of constants, $c^a$, is given by \be
c^a= \left(
\begin{array}{c}
d-4 \\
d-4 \\
d-4 \\ 
d-4 \\
d-4 \\
d-4 \\
2\, d -10
\end{array}
\right) ~.
\label{def:c}
\ee

The generator matrix $Tx^a_j \del^j $ is given by \be
Tx^a_j \del^j= -2\( \begin{array}{cccccc}
x_1	 & s^6_L	& 0	& 0		& s^2 	& 0 \\
s^6_L & x_3 	& 0	& 0 		& s^4	& 0 \\
 s^2	& s^4	& 0	& 0		& x_5	& 0 \\
 0 	& 0		& x_2 & s^6_R & s^1 	& 0 \\
 0	& 0 		& s^6_R & x_4 & s^3 	& 0 \\
 0 	& 0 		& s^1	& s^3 & x_5 	& 0 \\
 x_1	& x_3	& x_2 	& x_4 & x_5 	& x_6 \\
\end{array} \) 
\( \begin{array}{c}
\del^1 \\
\del^3 \\
\del^2 \\
 \del^4 \\
 \del^5 \\
 \del^6 \\
 \end{array} \) ~.
\label{def:Tx}
\ee
Note the change in order between $\del^2$ and $\del^3$ in order to highlight the block structure\footnote{
In hindsight we would have exchanged the initial labelling of 2, 3.}.  
The $s$ variables are defined as follows \bea
 s^1 := (x_5 + x_2 - x_1)/2 \qquad s^2 := (x_5 + x_1 - x_2)/2 \non
 s^3 := (x_5 + x_4 - x_3)/2 \qquad  s^4 := (x_5 + x_3 - x_4)/2 \non
 s^6_L := (x_1+x_3 - x_6)/2 \qquad  s^6_R := (x_2+x_4 - x_6)/2 
\label{def:s}
\eea
These definitions are inspired by the definition of the $s$ variables in the diameter diagram \cite{diam}, where they were defined to be Legendre conjugates of the $x$ variables with respect to $\lambda$, the Heron / K\"all\'en invariant (see e.g. \cite{SFI,bubble} and references therein) given by \be
\la := x^2 + y^2 + z^2 - 2\, x\, y- 2\, x\, z - 2\, y\, z
\label{def:lam} 
\ee
More generally, every trivalent vertex $v \in \{L, R, T, B \}$, which stand for left, right, top and bottom vertices, defines a $\la$ variable \be
 \la_v:=\la(x_a, x_b, x_c)
\label{def:lamv}
\ee
  where $a,b,c$ denote the three propagators attached to $v$:  $L=(136),\, R=(246),\, T=(125)$ and $B=(345)$. The general $s$ variables are defined by \be
  s^a_v = -\del \la(a,b,c)/4 \del x_a  =(x_b + x_c - x_a)/2 ~.
\label{def:s_gen}
  \ee
This general definition includes those of  $s^6_L,\, s^6_R$ in (\ref{def:s}).
 
\begin{figure}
\centering \noindent
\includegraphics[width=13cm]{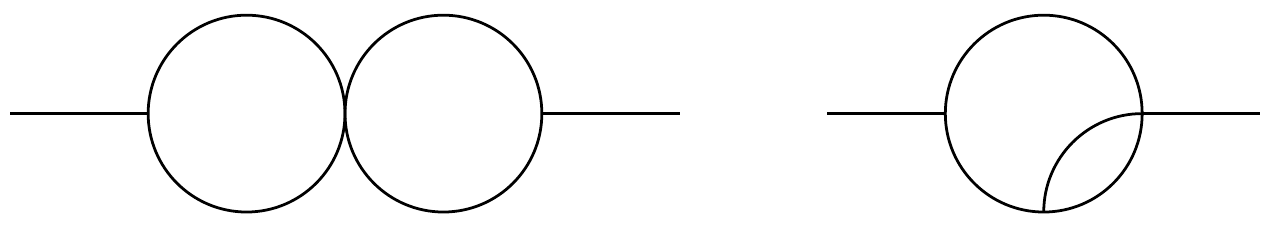} 
\caption[]{The two source topologies (a) figure 8 (b) propagator seagull (its vacuum closure is the vacuum seagull \cite{VacSeagull}).}
\label{fig:sources} \end{figure}

Finally the source vector $J^a$ is given by 
\be
J^a=\left(
\begin{array}{c}
 \del^5\, O_2 - (\del^3 + \del^5) O_1\\
 \del^5\, O_4 - (\del^1 + \del^5) O_3 \\
 \del^1\, O^2 + \del^3\, O^4 - (\del^1 + \del^3) O_5  \\
 \del^5\, O_4 - (\del^4 + \del^5) O_2 \\
 \del^5\, O_3 - (\del^2 + \del^5) O_4\\
  \del^2\, O^1 + \del^4\, O^3 - (\del^2 + \del^4) O_5\\
 0 \\
\end{array}
\right)
\label{def:J}
\ee
where the $O_i$ operators $i=1,\dots,5$ denote the diagram gotten by omitting, or contracting, the $i$'th propagator. Two possible topologies appear: figure 8 shown in fig. \ref{fig:sources}(a) and the propagator seagull shown in fig. \ref{fig:sources}(b).
This equation set was checked against the program FIRE \cite{Smirnov:2014hma}. 

%%%%%%%%%%%%%%% sources figures %%%%%%%%%%%%%%%

%%%%%%%%%%%%%%%%%%%%%%%%%%%%%%%%%%%%

The basis (\ref{def:basis}) is chosen in a way which is compatible with the discrete symmetry group $\Gamma$ (\ref{def:Gamma}). Specifically $E^1, E^2, E^4, E^5$ transform into each other under the various reflections. In this sense, it would have been sufficient to define $E^1$ and then the other three could have been defined by reflections. $E^3$  is invariant under up-down reflection, and a right-left reflection generates $E^6$. Another property is that the first three equations $E^1, E^2, E^3$ include derivatives with respect to the left part of the diagram, namely $\del^1, \del^3, \del^5$. Upon left-right reflection they transform into $E^4, E^5, E^6$ respectively. Finally $E^7$ is the dimension equation, namely Euler's identity for homogeneous functions. 

%%%%%%%%%%%%%%%%%%%%%%%%%%%%%%%%%%%%%%%%%%%
\section{Geometry of parameter space}
\label{sec:geom}

In this section we analyze the geometry in parameter space.

\presub {\bf $G$-orbit co-dimension and 6-minors}. The equation set (\ref{eq_set}) consists of 7 equations in a 6 dimensional parameter space. The dimension of the $G$-orbit through any point $x \in X$ is given by the rank of $Tx$ at that point. 

In order to determine the rank we follow the method of maximal minors \cite{minors} and compute the 6-minors $M_a$ defined by omitting row $a$ taking a determinant and multiplying by an alternating sign (see \cite{minors} for a precise definition in terms of the $\ep$ tensor). $M_a$ is found to be \be
M_a =4\, p^2\, B_3(x)\, K_a(x)
\label{6minor}
 \ee
 where the notation $B_3(x), K_a(x)$ will be defined now.
 
$B_3(x)$ is a cubic polynomial defined by \bea \label{def:B3}
B_3 &=&  x_1\, x_4 (x_1 + x_4) + x_2\, x_3 (x_2 + x_3) +  x_5\, x_6 (x_5 + x_6) +  \non
 &+&  x_1\, x_2\, x_5 + x_1\, x_3\, x_6 + x_2\, x_4\, x_6 +  x_3\, x_4\, x_5   \\
&-& [ x_1\, x_4 (x_2 + x_3 + x_5 + x_6) + x_2\, x_3 (x_1 + x_4 + x_5 + x_6) + x_5\, x_6 (x_1 + x_2 + x_3 + x_4)  ] \nonumber 
\eea
The first line includes a sum over the 3 opposing edge pairs of the diagram's vacuum closure which is a tetrahedron, the second line sums over its 4 vertices and the last line sums over the 12 collections of distinct edges which neither share a vertex nor form a face.\footnote{The first and third lines can be written alternatively as $2 x_1\, x_4 (x_1 + x_4) + 2 x_2\, x_3 (x_2 + x_3) +  2 x_5\, x_6 (x_5 + x_6) - (x_1 + x_2 + x_3 + x_4 + x_5 + x_6) (x_1 x_4 + x_2 x_3 + x_5 x_6) $.}

\begin{figure}
\centering \noindent
\includegraphics[width=13cm]{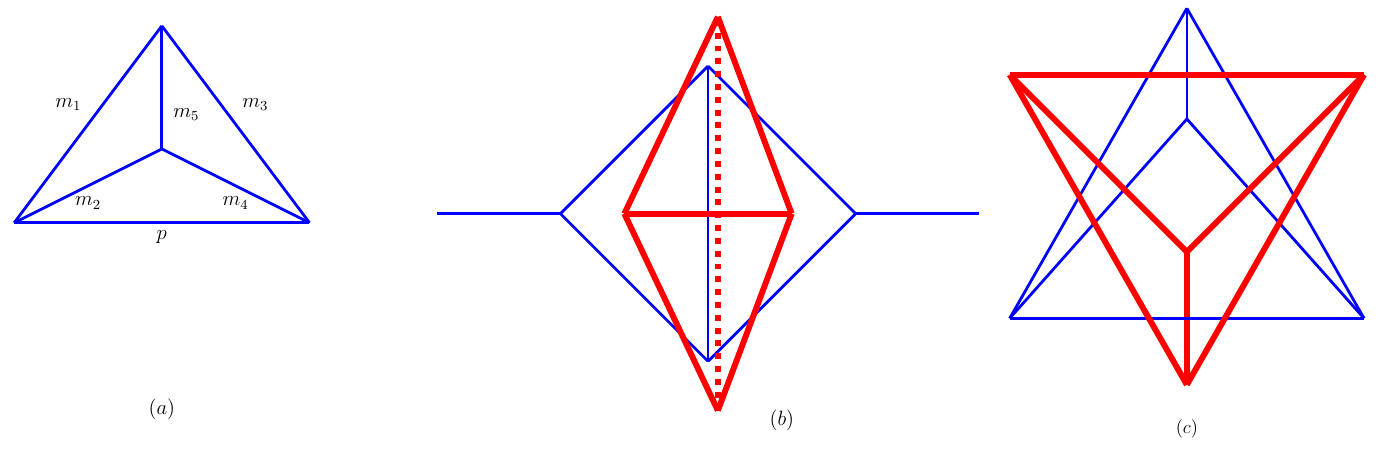} 
\caption[]{The dual tetrahedron which appears in the analysis of the kite diagram. (a) Top view with edge size indicated. (b)  The dual diagram (heavy red line) drawn on the same plane together with the Feynman diagram. The line dotted can be thought to be outside of the plane.
(c) 3d view of two dual tetrahedra. }
\label{fig:dual_tetra}  \end{figure}

According to the Cayley-Menger formula $B_3$ describes the squared volume of a tetrahedron \cite{wiki_Cayley_Menger}.  More specifically, it is the volume of the tetrahedron which is dual to the vacuum closure of the diagram -- see fig. \ref{fig:dual_tetra}: the duality exchanges vertices and faces, the edge dual to the incoming momenta is $p^\mu$ and each of the other edges is of length $m_i$, the mass of the dual propagator \cite{tetrahedron}. 

$B_3$ appeared in the physics literature in the work of Baikov on the 3-loop vacuum diagram (tetrahedron) \cite{Baikov1996a, Baikov1996b}. Therefore we shall refer to $B_3$ as the Cayley-Menger / Baikov polynomial. %, or the Baikov polynomial in short. In the notation $B_3$, $B$ stands for Baikov while $3$ reminds us that it is cubic. 
More generally, the Cayley-Menger formula specifies the squared-volume of the $n$-simplex through a determinant, so $\lambda$ and $B_3$ are specific instances of it.

The dual tetrahedron which appears here is familiar also from the formulation of Landau singularities \cite{Landau1959}.

The vector $K_a$ is given by \be
K = {\left( \begin{array}{c}
 -\partial^2\, B_{3} \\
-\partial^4\, B_{3} \\
 \lambda_L \\
 \partial^1\, B_{3}  \\
 \partial^3\, B_{3} \\
 -\lambda_R  \\
 0 \\ 
 \end{array} \right)}^{T}
\label{def:K}
\ee
$\la_L,\, \la_R$ are defined in (\ref{def:lam}-\ref{def:lamv}). They are related to derivatives of $B_3$ by the identity \be
\la_L = \( \del^2 + \del^4 + \del^5 \) B_3
\label{lam_id}
\ee
and similarly $\la_R=  \( \del^1 + \del^3 + \del^5 \) B_3$. These identities can be confirmed through direct calculation.

For convenience we mention the detailed expression for a derivative of $B_3$ \be
\del^1\, B_3 = x_4^2 + (x_2-x_6) (x_5-x_3)-x_4 ( x_2+x_3+x_5+x_6-2 x_1) ~.
\ee 
 All other derivatives $\del^i B_3$ can be obtained by permutations. 

$K_a$ is a global stabilizer, namely it satisfies \be
K_a\, Tx^a_j = 0
\ee
 everywhere in $X$ (so it leaves $x$ invariant) \cite{minors}. The existence of a single global stabilizer is to be expected since we have 7 equations, yet the dimension of the $G$ orbit is at most 6. Since $K_a\, c^a =0$ multiplying the equation set (\ref{eq_set}) by $K_a$ generates a global constraint among the sources $K_a\, J^a=0$. 

From (\ref{6minor}) we read that the common factor $S(x)$, termed the singular locus polynomial \cite{minors} is \be
S(x) = p^2\, B_3 ~.
\label{def:Sing}
\ee

For generic values of $x \in X$ $S(x) \neq 0$ and $M_a(X) \neq 0$ and hence the dimension of the $G$-orbit is generically 6. We confirmed this by a numerical evaluation of ${\rm rk} (Tx)$ at randomly chosen points. Since ${\rm dim}(X)=6$ we may answer the first question from the introduction and conclude that generically in $X$ \be
 {\rm codim} (G-{\rm orbit}) = 0 ~. 
\label{G-orb}
\ee
This means that SFI is maximally effective for the kite diagram and a discrete set of base points in $X$ space will suffice for reaching any other point through a line integral over a path which lies within a $G$-orbit.

\presub {\bf Homogeneous solution}. The homogeneous solution of the equation set (\ref{eq_set}), $I_0$ is an ingredient of the general reduction formula to a line integral. With that objective in mind we proceed to determine $I_0$. 

The constant free subgroup of $G$ are defined here to be linear combinations of equations
with $d$-independent coefficients such that the constant term vanishes. Hence, by definition $I_0$ is annihilated by $G_{cf}$ and it must be a function of $G_{cf}$ invariants.

We have ${\rm dim}(G_{cf})=5$ since the rank of the components\footnote{
w.r.t. $d$-independent coefficients, of course.}
 of the constant vector $c^a$ is 2. We choose a basis for the constant free equations $F^1,\dots,F^5$ as follows \be
 \( \begin{array}{c}
	F^1 \\ 
	F^2 \\ 
	F^3 \\ 
	F^4 \\ 
	F^5 \\ 
	\end{array} \) =
	 \( \begin{array}{c}
	E^1-E^3 \\ 
	E^2-E^3 \\ 
	E^4 - E^6 \\ 
	E^5 - E^6 \\
	E^3 - E^6 \\ 
	\end{array}
	\) 
\label{def:F}
\ee
The generator matrix reads \be
Tx_{cf} = -2\( \begin{array}{cccccc}
s^5_T & s^6_L-s^4	& 0	& 0			& -s^1 	& 0 \\
s^6_L-s^2 & s^5_B 	& 0	& 0 			& -s^3	& 0 \\
 0 	& 0			& s^5_T & s^6_R-s^3 & s^1 	& 0 \\
 0	& 0 			& s^6_R-s^1 & s^5_B & s^3 	& 0 \\
 s^2	& s^4		& -s^1	& -s^3 & 0 	& 0 \\
 \end{array} \)
\label{Tx_cf}
\ee
where the $s$ variables were defined in (\ref{def:s}) and the definitions of $s^5_T, s^5_B$ follow the same notation while referring to the top and bottom vertices, explicitly \be
s^5_T := (x_1+x_2 - x_5)/2 \qquad  s^5_B := (x_3+x_4 - x_5)/2
\label{def:s5}
\ee

This basis for $G_{cf}$ was chosen once again to be $\Gamma$ compatible in the sense that $F^1$ can generate $F^2, F^3, F^4$ under reflections, while $F^5$ is a singlet (even under up-down reflection and odd under right-left reflection).

The global stabilizer $K_a$ (\ref{def:K}) is within $G_{cf}$ (since $K_a c^a=0$). Hence point-wise in $X$ the 5 generators of $G_{cf}$ have a single relation and the dimension of the orbits is \be
{\rm dim } (G_{cf}-{\rm orbit}) = 4
\ee
Therefore $G_{cf}$ has two independent invariants.

The invariants of $G_{cf}$ turn out to be $p^2, B_3$, namely $(Tx_{cf})^a_j \del^j p^2=(Tx_{cf})^a_j \del^j B_3 =0$ which is confirmed through a straightforward computation. The form of the invariants can be motivated as follows. Since the 6th row of the generator matrix (\ref{Tx_cf}) vanishes one recognizes that $x_6 \equiv p^2$ would be annihilated by it. Next, since $G_{cf} \subset SL(3,\IR)$, $G_{cf}$ preserves volume in the space of currents and hence $B_3$ which represents a volume would be expected to be preserved. 

A more systematic derivation of the invariants through the method of maximal minors is offered in appendix \ref{app:invar}.

Substituting $I_0=I_0(p^2, B_3)$ into the equation set (with $J^a$ put to zero) we obtain the set \bea
 (d-4)I_{0}-2 B_{3}\frac{\partial I_{0}}{\partial B_{3}} &=& 0 \non
 (d-5)I_{0}-3 B_{3}\frac{\partial I_{0}}{\partial B_{3}}-p^2 \frac{\partial I_{0}}{\partial p^2} &=& 0  ~.
 \label{I0_set}
 \eea
The first equation is gotten by substitution into any of the equations $E_1,\dots,E_6$ while the second originates in $E_7$, the dimension equation. The solution to this equation set provides us with the homogenous solution $I_0$ 
\be
I_0 =  {p^2}^{(1-\frac{d}{2})} B_3^{\frac{d-4}{2}} ~.
\label{I0_expr}
\ee 

The remaining steps to obtain the reduction to a line integral will be discussed within the open questions part of the last section.

%%%%%%%%%%%%%%%%%%%%%%%%%%%%%%%%%%%%%%%
\section{Algebraic locus and solutions}
\label{sec:algeb}

At the singular locus, namely when $B_3(x)=0$ or $p^2=0$ the dimension of the $G$ orbit is reduced and accordingly an additional stabilizer appears. Given a stabilizer $Stb_a$, if the associated constant is non-zero, namely $Stb_a\, c^a \neq 0$ one can reduce the diagram to a linear combination of simpler ones by multiplying the equation set on the left by the stabilizer. In such a case the set of differential equations degenerates into an algebraic equation, the associated component of the singular locus is called an algebraic locus and the resulting expression for the diagram is called the algebraic solution.

In this section we obtain the algebraic solution for the diagram at the $B_3(x)=0$ algebraic locus (the $B_3$ locus in short), and provide the stabilizers for the $p^2$ locus.

\presub {\bf $B_3$ locus}.  At $B_3=0$ the global stabilizer $K$ (\ref{def:K}) splits into a pair of stabilizers $K^L,\, K^R$ as follows \be
K^L = {\left( \begin{array}{c}
 -\partial^2\, B_{3} \\
-\partial^4\, B_{3} \\
 \lambda_L \\
 0  \\
 0 \\
 0  \\
 0 \\ 
 \end{array} \right)}^{T} 
 \qquad 
 K^R = {\left( \begin{array}{c}
0 \\
0 \\
 0 \\
 -\partial^1\, B_{3}  \\
- \partial^3\, B_{3} \\
 \lambda_R  \\
 0 \\ 
 \end{array} \right)}^{T} 
\label{def:KLR}
\ee

It is immediate to confirm that they are indeed stabilizers since $K^R_a Tx^a_j \del^j=K^L_a Tx^a_j \del^j=-2 B_3(x) \del^5 =0 ~ ({\rm mod}\, B_3)$.  $K^L$ is called the left stabilizer since only its first 3 component are non-zero, those which multiply first 3 equations in our basis (\ref{def:basis}), namely, those which involve only derivatives w.r.t. the left propagators $1,3,5$ and similarly for $K^R$. We note that $K=K^L-K^R$ and hence the global stabilizer (\ref{def:K}) is within the span of $K^L,\, K^R$ as it must, since the whole stabilizer space is 2-dimensional.

The 3-vector appearing in $K^L$ has the following alternate forms mod $B_3$ and up to overall scale \be
u^L = {\left( \begin{array}{c}
 -\partial^2 \\
-\partial^4 \\
 \del^2+\del^4+\del^5 \\
 \end{array} \right)} B_3,~
 v^L = {\left( \begin{array}{c}
 -\del^6  \\
 \del^3+\del^4 + \del^6  \\
 -\del^4 \\
 \end{array} \right)} B_3,~
 w^L = {\left( \begin{array}{c}
 \del^1+\del^2+\del^6 \\
-\del^6 \\
 -\del^2 \\
 \end{array} \right)}B_3
 \ee
$u^L$ consists of the 3 non-zero components of $K^L$. While $u^L,\, v^L,\, w^L$ appear altogether different, they are in fact all parallel at the $B_3$ locus as confirmed by computing the cross product. For example \be
u^L \times v^L = 4\, B_3 (x_1,\, s^6_L,\, s^2_T) ~.
\label{uv_cross}
\ee
Either $u^L,\, v^L$ or $w^L$ on its own would not have provided a complete description of the stabilizer since each vanishes on some 2d manifold. Together they provide alternate sections of the same line bundle. 

Similarly the 3-vector which appears in $K^R$ has the following alternate forms \be
u^R = {\left( \begin{array}{c}
 -\partial^1 \\
-\partial^3 \\
 \del^1+\del^3+\del^5 \\
 \end{array} \right)} B_3,~
 v^R = {\left( \begin{array}{c}
 -\del^6  \\
 \del^3+\del^4 + \del^6  \\
 -\del^3 \\
 \end{array} \right)}  B_3,~
 w^R = {\left( \begin{array}{c}
 \del^1+\del^2+\del^6 \\
-\del^6 \\
 -\del^1 \\
 \end{array} \right)} B_3
 \ee
These 3-vectors are related by reflections: left-right reflection exchanges $R \leftrightarrow L$ while up-down reflection exchanges $v \leftrightarrow w$.
 
The additional stabilizer was derived by solving $B_3=0$ for one of the variables (we chose $x_6$; $B_3$ is quadratic in it), substituting back into the matrix $Tx$, solving for the right null vectors through standard Gauss elimination, then finally restoring $x^6$ to eliminate square roots. In this way we obtained $v^L,\, u^R$ and the rest were obtained through symmetry operations. It would be interesting to obtain the stabilizer 2-form $K_{ab}$ through the method of maximal minors.
 
\presub {\bf Algebraic solution}. The algebraic solution is now gotten by multiplying the equation set $(\ref{eq_set})$ on the left by an arbitrary linear combination $\al_L K^L + \al_R\, K^R$. We notice that \be
\( \al_L K^L_a+ \al_R\, K^R_a \)\, c^a =  (\al_L+\al_R)\, (d-4)\, \del^5 B_3
\label{def:c_K}
\ee
where $c^a$ is given at (\ref{def:c}) and the identity (\ref{lam_id}) was used. The algebraic solution is given by the following alternate forms \begin{subequations} \label{alg_soln}
\begin{align}
 (4-d)\, I &= \frac{ \( \al_L K^L_a+ \al_R\, K^R_a \) \, J^a} {(\al_L+\al_R)\, \del^5 B_3} =
  \label{alg_geom_a} \\
  &= \frac{u^L \cdot J_L}{\del^5 B_3} =  \frac{u^R \cdot J_R}{\del^5 B_3} = 
  \label{alg_geom_b} \\
  &= \frac{v^L \cdot J_L}{\del^3 B_3} = \frac{w^L \cdot J_L}{\del^1 B_3} = \frac{v^R \cdot J_R}{\del^4 B_3} = \frac{w^R \cdot J_R}{\del^2 B_3}
  \label{alg_geom_c} 
  \end{align}
\end{subequations}
where in the first line the sources $J^a$ are given in (\ref{def:J}) in terms of simpler diagrams, and $K^L, K^R$ are defined in (\ref{def:KLR}). In the second and third lines we defined the left and right source 3-vectors $J_L,\, J_R$ \be
J_L = {\left( \begin{array}{c}
 J^1 \\
J^2 \\
 J^3 \\
 \end{array} \right)} 
 \qquad
 J_R = {\left( \begin{array}{c}
 J^4  \\
 J^5  \\
 J^6 \\
 \end{array} \right)}  
 \label{def:Ji}
 \ee
In the second line we put either $\al_R=0$ or $\al_L=0$ to get alternative expressions related by left-right reflection, and in the third line we use the equivalence of $u,v,w$ to get four more alternative expressions which are related to each other through reflections. 
This answers the second question from the introduction and it is our main result.

The algebraic solution must be independent of the choice of $\al_L,\, \al_R$. For diagrams which are left-right symmetric, namely $m_1=m_2,\, m_3=m_4$ this is apparent, since the numerator also becomes proportional to $(\al_L+\al_R)$. In addition all four forms on (\ref{alg_geom_c}) must be equivalent. A general demonstration of this independence and equivalence of forms appears to require knowledge of relations among the propagator seagull source diagrams. 

Tests and special cases. In the massless case $m_1=\dots=m_5=0$ it was shown already in \cite{ChetyrkinTkachov1981} that the diagram can be reduced as follows \be
I_{\rm massless} = \frac{2}{4-d} \( {\rm fig.}\; \ref{fig:sources}(b)'- {\rm fig.}\;  \ref{fig:sources}(a)' \)
\ee
The primes denote that the top propagator of (b) (propagator seagull) should be squared and so should one of the propagators in (a) (figure 8 diagram).  This identity was originally provided as a simple example for the Integration By Parts (IBP) method, when the latter was introduced. Interestingly it was used later in an essential way in the computation of the two-body effective potential for the binary problem at the second post-Newtonian order (2PN) \cite{GilmoreRoss2008}.  
% There it is given by the equiv. expression 
% 	2/(d-4) [ eight  - prop.seagull ]
The restriction of  (\ref{alg_geom_a}) to the massless case is independent of $\al^L, \al^R$ and we find full agreement with this expression. In this case the forms in (\ref{alg_geom_c}) are all of the form $0/0$ and hence ill-defined, at least at face value.
 
The result  (\ref{alg_soln}) generalizes the reduction of the massless case to the most general parameters, namely $B_3(m_1^2, \dots, m_5^2, p^2)=0$.
 
The case $m_3=m_4=m_5=0$ is of special interest. In this case $B_3=0$ and  (\ref{alg_soln}) simplifies to  the following two alternative forms \be
(4-d) I_{x_3=x_4=x_5=0} =  \frac{(x_2-x_1)J^2 + (x_1-x_6) J^3}{x_2-x_6}= \frac{(x_1-x_2) J^5 + (x_2-x_6) J^6}{x_1-x_6} ~,
\ee
where  the source components $J^i$ were defined in (\ref{def:J}). This case falls into the applicability regime of the ``diamond rule'' \cite{Ruijl:2015aca} (with $L=S=1$).\footnote{We thank K. Chetyrkin for this observation.} It would be interesting to extract a concrete expression from that approach and compare with the expressions here.
 
\presub {\bf $p^2$ locus}. In this case we too we find a pair of stabilizers, they are given by \be
t^L = {\left( \begin{array}{c}
 -2 x_3 \\
2 x_1 \\
 0 \\
 -2s^5_B  \\
 2 s^5_T \\
 x_4 - x_2  \\
 x_3 - x_1 \\ 
 \end{array} \right)}^{T} 
 \qquad 
 t^R = {\left( \begin{array}{c}
-2 s^5_B \\
2 s^5_T \\
 x_3 - x_1 \\
 -2 x_4  \\
 2 x_2 \\
 0  \\
 x_4 - x_2 \\ 
 \end{array} \right)}^{T} 
\label{def:tLR}
\ee
Each one is odd under up-down reflections and they are exchanged by left-right reflection. The global stabilizer is a linear combination given by \be
\left. K \right|_{p^2=0} = (x_2-x_4)\, t^L + (x_3-x_1)\, t^R
\ee
  
These stabilizers can be used to express the kite with $p^2=0$ in terms of simpler diagrams. In fact, it is expected to be described by the considerably simpler diameter diagrams, and hence we did not pursue it in this paper.

%%%%%%%%%%%%%%%%%%%%%%%%%%%%%%%%
\section{Summary and discussion}
\label{sec:summ}

In this paper we have explored the kite diagram through the Symmetries of Feynman Integral method (SFI). We were able to answer the questions in the introduction, as follows \bi

\item The $G$-orbits were found to be 6-dimensional in our 6d parameter space $X$, namely the orbit co-dimension is zero (\ref{G-orb}). This means that for this diagram the SFI method would be maximally effective. 

\item On the surface $B_3=0$ where $B_3=B_3(\{ x_i\})$ is given by  (\ref{def:B3}),  the integral degenerates into a linear combination of simpler diagrams and is given by (\ref{alg_soln}), thereby providing a maximal generalization of the massless case. This is our central result. 

\ei

We are not familiar with other studies of the kite diagram with most general parameters, and in particular the above-mentioned expressions for the algebraic locus and the algebraic solution on it.

\presub {\bf Open Questions}. We leave a few question for further study:

Simplification of source. It would be interesting to be able to simplify the sources such that the global constraint would simplify to zero. This should allow a simpler form for both the algebraic solution and the reduction to a line integral. However, this appears to require new relations for the propagator seagull diagram shown in fig. \ref{fig:sources}(b).

Base point. It would be interesting to determine the value of the diagram at some point $x$ where $B_3(x) \neq 0$ and hence can serve as a base point for the line integral representation throughout the parameter space $X$. In particular one could choose $x_1=x_2=x_3=x_4=0$ so that only the middle mass is non-zero. In addition one would need to specify a choice of a path for the line integral.
 
 \presub{\bf Discussion}.
 
It would be interesting to test our results through numerical evaluation. In particular one could numerically evaluate both sides of (\ref{alg_soln}) through parametric integration and compare.

We would like to make a general comment about integrals with non-unit indices (namely, powers of propagators $\nu_i>1$). As discussed in \cite{SFI} if the dependence of the integral on all possible parameters is known, then higher indices can be obtained through derivatives. Moreover, while non-unit indices are surely of interest, in some sense, unit indices are more natural and frequent. Now, it should be noted that as long as the dependence on the parameters is not known in full, but rather only in some sector, then to obtain non-unit indices  may require to combine our method with the more standard IBP reduction to master integrals, such as in \cite{Tarasov:1997kx}.

%%%%%%%%%%%%%%%%%%%%%%%%%%%%%%%%%%%%%%%%% 
\subsection*{Acknowledgments}
 
 We would like to thank Ruth Shir and Amit Schiller for many useful discussions. S. M. would like to thank ICTP, Trieste and TIFR, Mumbai for hospitality while this work was in progress.
 
This research was  supported by the ``Quantum Universe'' I-CORE program of the Israeli Planning and Budgeting Committee.

%%%%%%%%%%%%%%%%%%%%%%%%%%%%%%%%%%%%%%%%%%%%%%%%%%% 

\appendix

\section{Invariants of the constant free subgroup}
\label{app:invar}

It is straightforward to confirm that $p^2$ and $B_3$ (\ref{def:B3}) are invariants of the constant free subgroup $G_{cf}$ (\ref{def:F}). This section describes a systematic derivation of the invariants through the method of maximal minors \cite{minors}.

The matrix $Tx_{cf}$ (\ref{Tx_cf}) has 5 rows (equations) and 6 columns (variables in parameter space). Hence we start by considering the 5-minors. Since the 6th row consists of zeros we conclude that $M^i$ for $i=1,\dots,5$. By direction computation one finds that $M^6=0$ as well. Hence altogether \be
M^i=0 \qquad i=1,\dots,6
\ee
namely, the generic dimension of $G_{cf}$ orbit is less than 5, and we proceed to consider 4-minors.

Computation of 4-minors confirms that they are not identically zero, and hence generically \be
{\rm dim }(G_{cf}-{\rm orbit}) = 4 ~.
\ee
Moreover,one finds the factorization expected of maximal minors \be
M^{ij}_{~a} = Inv^{ij} K^{cf}_a
\ee
Here the constant-free stabilizer $K^{cf}$ is given by \be
K^{cf} = {\left( \begin{array}{c}
 -\partial^2\, B_{3} \\ % K^1
-\partial^4\, B_{3} \\ % K^2
 \partial^1\, B_{3}  \\ % K^4 
 \partial^3\, B_{3} \\  % K^5
 \del^5 B_3	 \\ 
 \end{array} \right)}^{T}
\label{def:K_cf}
\ee
It is the same stabilizer as the previously found global stabilizer $K$ (\ref{def:K}) only transformed to a different basis, since indeed $K$ is constant-free, namely $K_a c^a = 0$.

The invariant tensor $I^{ij}$ is found to be of the form \be
I^{ij} = \del^{[i} B_3 \delta^{j]}_6 = dB_3 \wedge dp^2
\ee
and hence we deduce that both $B_3$ and $p^2$ are invariants of $G_{cf}$ (in the last equation we did not need to be careful about signs).

\bibliographystyle{JHEP}
\bibliography{kite_bib}

\end{document}